\def\dsp{\displaystyle}
\def\oneh{{\textstyle {1\over 2}}}
\newcommand{\tvec}[1]{\mbox{\boldmath{$#1$}}}

\documentclass[
    ,final            
  ]
  {aipproc}

\layoutstyle{6x9}

\begin{document}

\title{Structure of the Nucleon Spin on the Light Cone}


\author{B. Pasquini}{
  address={
Dipartimento di Fisica Nucleare e Teorica, Universit\`{a} degli Studi di Pavia, and\\
Istituto Nazionale di Fisica Nucleare, Sezione di Pavia, I-27100 Pavia, Italy}
}



\begin{abstract}
The spin structure of the nucleon is studied
in a light-cone description of the nucleon where the Fock expansion is 
truncated to consider only valence quarks.
Transverse momentum dependent parton distributions and 
transverse-spin densities, defined through the generalized parton distributions
in the impact parameter space, are investigated as new tools to reveal
the spin-spin and spin-orbit correlations for different quark and nucleon 
polarizations.
\end{abstract}

\maketitle


\section{Introduction}

One of the most important goal in QCD spin physics is to understand the spin structure of the nucleon, i.e. how the nucleon spin is made from its fundamental 
constituents.
This issue has been of intensive experimental and theoretical investigation 
over the last decades, in particular by exploring the QCD parton model
in deep inelastic processes in terms of unpolarized  and helicity parton distributions functions (PDFs).
In recent years
 a precise knowledge of the transverse structure as well as of 
parton-momentum
correlations  has emerged as an essential part to unravel the spin and momentum substructure of the nucleon.
This information can only be obtained by considering processes beyond the 
inclusive reactions, such as semi-inclusive deep-inelastic (SIDIS) lepton nucleon scattering or exclusive lepto-production processes 
at large momentum transfer.
According to the factorization theorem,  
the physical observables of such processes can be expressed as convolution of 
hard partonic scattering cross sections, and soft non-perturbative
objects given by quark-quark correlation functions
which 
generalize the forward matrix elements occurring in the definition of
the PDFs.
In the case of SIDIS processes 
 one deals with
 transverse momentum 
dependent parton distributions (TMDs) which 
are defined in terms of the same matrix elements entering the definition of 
PDFs, but without integration over the transverse momentum~\cite{Mulders2}.
On the other hand,
exclusive processes such as $e N\rightarrow e N\gamma$ or
$e N\rightarrow e N M$ (where $M=\pi,\,\rho, \,K$ etc.) 
allow to access
generalized parton distributions (GPDs) which involve non-forward matrix elements 
between hadron states with different momentum and/or polarization before and after the scattering~\cite{Jiarnps}. 
Although the TMDs and GPDs  can be seen as
 two different limiting cases
of the same generalized parton correlation functions, 
no-model independent relations between the two classes of objects has been 
obtained so far~\cite{Meissner}.
\newline
\noindent
The TMDs contain rich and direct three-dimensional information about the 
internal dynamics of the nucleon, and in particular can help in understanding 
the strength of different spin-spin and spin-orbit correlations.
On the other hand, the GPDs provide new method of spatial imaging of the 
nucleon, through the definition of impact-parameter dependent spin
densities  which reveal the correlations between the momentum and spatial 
distributions of quarks for different quark and target polarizations.
A convenient way to make explicit which kind of information on hadron 
structure is contained in these quantities is the representation in terms 
of overlap 
of light-cone wave functions (LCWFs) which are the probability amplitudes
to find a given $N$-parton configuration in the Fock-space expansion 
of the hadron state~\cite{BPP98}.
In the following, we will confine our analysis to the three-quark sector, 
by truncating the light-cone expansion of the nucleon state to the minimal
Fock-space configuration.
The three-quark component of the nucleon have been studied 
extensively in the literature~\cite{LB80}-\cite{Stefanis} in terms of quark distribution 
amplitudes defined as hadron-to-vacuum 
transition matrix elements of non-local gauge-invariant light-cone operators.
Unlike these works,
the authors of Refs.~\cite{BurkJiY02,Ji:2002xn} considered the 
wave-function amplitudes keeping full transverse-momentum dependence of 
partons and proposed a systematic way to enumerate independent amplitudes of 
a LCWF which parametrize the different orbital angular momentum components
of the nucleon state.
This approach consists in writing down the  matrix elements of a class 
of three-quark
light-cone quark operators which serve to define a complete set of 
light-cone amplitudes.  These matrix elements can
 be simplified using color, flavor, spin and discrete 
symmetries~\cite{Ji:2002xn}, and at the end one finds that six amplitudes are needed to describe the three-quark sector of the nucleon LCWF.
This general classification scheme can be  used to obtain the overlap LCWF 
representation of the TMDs and GPDs, 
which in turn can be applied 
to obtain predictions within specific dynamical models of the nucleon.
Here
we will adopt 
a light-cone constituent quark model (CQM) which has been successfully 
applied in the calculation of the electroweak properties of the nucleon~\cite{PBff}.
As outlined in Ref.~\cite{BPT1},
the starting point is the three-quark wave function obtained as solution
of the  Schr\"odinger-like eigenvalue equation in the instant-form dynamics.
The corresponding solution in light-cone dynamics is obtained through the
unitary  transformation represented by product of Melosh rotations acting
on the spin of the individual quarks.
In particular, the instant-form wave function is constructed as a product of a momentum wave function which is spherically symmetric and invariant under permutations, and a spin-isospin wave function which is uniquely determined by SU(6)
symmetry requirements.
By applying the Melosh rotations, the Pauli spinors of the quarks in the
 nucleon rest frame are converted to the light-cone spinor.
The relativistic spin effects are evident in the presence
of spin-flip terms in the Melosh rotations which generate non-zero orbital angular momentum component
and non-trivial correlations between spin and transverse momentum of the 
quarks. On the other hand,
the momentum-dependent wave function keeps the original functional form, with 
instant-form coordinates rewritten in terms of light-cone coordinates.
The explicit expressions of the light-cone amplitudes within this CQM 
 can be found in Ref.~\cite{PCB}, while the corresponding results for the TMDs and GPDs in the impact-parameter space will be discussed in 
sect. 2 and 3, respectively.

\section{Transverse-momentum dependent distributions}
SIDIS processes are described at leading order by eight TMDs,
$f_1^q,$ $f_{1T}^{\perp \, q},$ $g_1^q,$  $g_{1T}^q,$ $g_{1L}^{\perp\, q},$
$h_1^q,$  $h_{1T}^{\perp\, q},$ $h_{1L}^{\perp\, q}$, and 
$h_{1}^{\perp\, q},$ which
depend on $x$ and ${\bf k}^2_\perp.$ Among them,
 the Boer-Mulders $h_1^\perp$~\cite{BoerMulders} and the Sivers $f_{1T}^\perp$~\cite{Sivers} functions are T-odd, 
i.e. they change sign under ``naive time reversal'', 
which is defined as usual time reversal, but without interchange of initial 
and final states.
Restricting the analysis 
to the remaining six T-even TMDs, we find 
the following results within our light-cone CQM~\cite{PCB} 
\begin{eqnarray}
\label{eq:f1}
f^q_1(x,{\bf k}^2_\perp)&=&
N^q
\int {\rm d}[1] {\rm d}[2]{\rm d}[3]
\delta(k-k_3)
\vert \psi(\{x_i\},\{{\bf k}_{\perp\,i}\})\vert^2,\nonumber
\\
g^q_{1L}(x,{\bf k}^2_\perp)&=&
P^q
\int {\rm d}[1] {\rm d}[2]{\rm d}[3]
\delta(k-k_3)
\vert \psi(\{x_i\},\{{\bf k}_{\perp\,i}\})\vert^2
\frac{(m+ x M_0)^2 -{\bf k}^2_{\perp}}{(m+ xM_0)^2 + {\bf k}^2_{\perp}},\nonumber\\\nonumber
\label{eq:g1}&&\\
g^{q}_{1T}(x,{\bf k}^2_\perp)&=&
P^q
\int {\rm d}[1] {\rm d}[2]{\rm d}[3]
\delta(k-k_3)
\vert \psi(\{x_i\},\{{\bf k}_{\perp\,i}\})\vert^2
\frac{2M(m+ xM_0)}{(m+ xM_0)^2 + {\bf k}^2_{\perp}}, \nonumber\\
&&
\label{eq:g1T}\nonumber\\
h^q_1(x,{\bf k}^2_\perp)&=&
P^q
\int {\rm d}[1] {\rm d}[2]{\rm d}[3]
\delta(k-k_3)
\vert \psi(\{x_i\},\{{\bf k}_{\perp\,i}\})\vert^2
\frac{(m+ xM_0)^2}{(m+ xM_0)^2 + {\bf k}^2_{\perp}},\nonumber\\
&&
\label{eq:h1}\nonumber\\
h^{\perp\,q}_{1T}(x,{\bf k}^2_\perp)&=&-
P^q
\int {\rm d}[1] {\rm d}[2]{\rm d}[3]
\delta(k-k_3)
\vert \psi(\{x_i\},\{{\bf k}_{\perp\,i}\})\vert^2
\frac{2M^2}{(m+ xM_0)^2 + {\bf k}^2_{\perp}},\nonumber\\
\label{eq:h1T}&&\nonumber\\
h^{\perp\, q}_{1L}(x,{\bf k}^2_\perp)&=&
- P^q
\int {\rm d}[1] {\rm d}[2]{\rm d}[3]
\delta(k-k_3)
\vert \psi(\{x_i\},\{{\bf k}_{\perp\,i}\})\vert^2
\frac{2M(m+ xM_0)}{(m+ xM_0)^2 + {\bf k}^2_{\perp}},\nonumber\\
&&
\label{eq:h1L}
\end{eqnarray}
where 
$\delta(k-k_3)= \delta(x-x_3)\delta({\bf k}_{\perp}-{\bf k}_{\perp\,3})$, and the integration measures are defined as
\begin{equation}
{\rm d}[1] {\rm d}[2]{\rm d}[3]=
{\rm d}x_1{\rm d}x_2{\rm d}x_3
\delta\left(1-\sum_{i=1}^3 x_i\right)
\frac{{\rm d}^2 {\bf k}_{\perp\,1}{\rm d}^2{\bf k}_{\perp\,2}
{\rm d}^2{\bf k}_{\perp\,3}}{[2(2\pi^3)]^2}
\delta\left(\sum_{i=1}^3 {\bf k}_{\perp\,i}\right).
\end{equation}
In Eqs.~(\ref{eq:h1L}),
the flavor dependence is given by the factors $P^u=\displaystyle{\frac{4}{3}}$,  $P^d=\displaystyle{-\frac{1}{3}}$, $N^u=2$ and $N^d=1,$  as
dictated by SU(6) symmetry.
A further consequence of the assumed SU(6) symmetry is the factorization
 in Eqs.~(\ref{eq:h1L}) of the momentum-dependent wave function
$\psi(\{x_i\},\{{\bf k}_{\perp\,i}\})$  
from the spin-factor arising from the Melosh rotations.
Thanks to this factorized form one finds the following relations
\begin{eqnarray}
\label{eq:61}
2h^q_1(x,{\bf k}^2_\perp)
&=&g^q_{1L}(x,{\bf k}^2_\perp)+\frac{P^q}{N^q}f^q_1(x,{\bf k}^2_\perp),\\
\frac{P^q}{N^q}f^q_1(x,{\bf k}^2_\perp)
&=&h_1^q(x,{\bf k}^2_\perp) -\frac{{\bf k}^2_\perp}{2M^2}h_{1T}^{\perp \,q}(x,{\bf k}^2_\perp),
\label{eq:61a}\\
h_{1L}^{\perp q}(x,{\bf k}^2_\perp)
&=&-g_{1T}^q(x,{\bf k}^2_\perp).
\label{eq:61b}
\end{eqnarray}
Eq.~(\ref{eq:61}) is a generalization of analogous relations discussed in 
Ref.~\cite{PPB,PPB07} and was also rederived  together with 
Eq.~(\ref{eq:61a}) in Ref.~\cite{Avakian08}.
Eq.~(\ref{eq:61b}) was already found in the diquark spectator model 
of Ref.~\cite{Mulders3}. In QCD TMDs should be all independent of each other.
 The limitation to three valence quarks implies that out of the six TMDs 
$f_1$, $g_{1L}$, $g_{1T}$, $h_1$,  $h_{1T}^\perp$, $h_{1L}^\perp$ only 
three are linearly independent. 
A similar situation occurs with the bag model~\cite{Avakian08}. 
In the diquark spectator model of Ref.~\cite{Mulders3}
the relations (\ref{eq:61}) and 
(\ref{eq:61a}) hold only for the separate scalar and axial contributions, 
while Eq.~(\ref{eq:61b}) is verified more generally for both $u$ and $d$ 
flavors. However, this is no longer true by considering different versions of
the diquark spectator model as in Refs.~\cite{Conti,Gamberg}.
Furthermore, combining 
the relations (\ref{eq:61}) and (\ref{eq:61a}), one finds 
for the
$h_{1T}^\perp$ distribution
 \begin{equation}
\frac{{\bf k}_{\perp}^2}{2M^2}\,h^{\perp\,q}_{1T}(x,{\bf k}^2_\perp) 
= g^q_{1L}(x,{\bf k}^2_\perp) - h^q_1(x,{\bf k}^2_\perp).
\label{eq:pretz}
\end{equation}
This result was already found in Ref.~\cite{Avakian08}. 
Integrating out transverse momenta and going to the non-relativistic limit 
where helicity and transversity distributions coincide, one finds that the 
first moment of $h_{1T}^\perp$ vanish identically. 
Thus, relation (\ref{eq:pretz}) supports the statement that $h_{1T}^\perp$ is 
a measure of relativistic effects. Relativity, responsible for a chiral-odd  
transversity distribution differing from a chiral-even helicity distribution, 
exhibits the chirally odd nature of  $h_{1T}^\perp$.
This is confirmed by the following relation that is also satisfied within our
 model:
\begin{eqnarray}
\label{eq:tmd_gpd}
h_{1T}^{(0)\perp\,q}(x)=\frac{3}{(1-x)^2}\tilde H_T^q(x,0,0),
\end{eqnarray}
where $\tilde H_T^q(x,0,0)$ is the forward limit of a chiral-odd generalized 
parton distribution  occurring in the case of parton and nucleon 
helicity flip (see, e.g., Refs.~\cite{PPB,BPreview})
and 
the transverse moments of $h_{1T}^{\perp\,q}$ are defined as
$h_{1T}^{(n)\perp q}(x)=\int d^2 k_\perp \left(\frac{{\bf k}_{\perp}^2}{2M^2}\right)^n \, h_{1T}^{\perp\,q}(x,{\bf k}^2_\perp)$. 
Eq.~(\ref{eq:tmd_gpd}) was first found in Ref.~\cite{Meissner} 
to hold for the scalar diquark model and in a quark target model 
of the nucleon.
\newline
\noindent 
\begin{figure}
  \includegraphics[width=11.5 truecm]{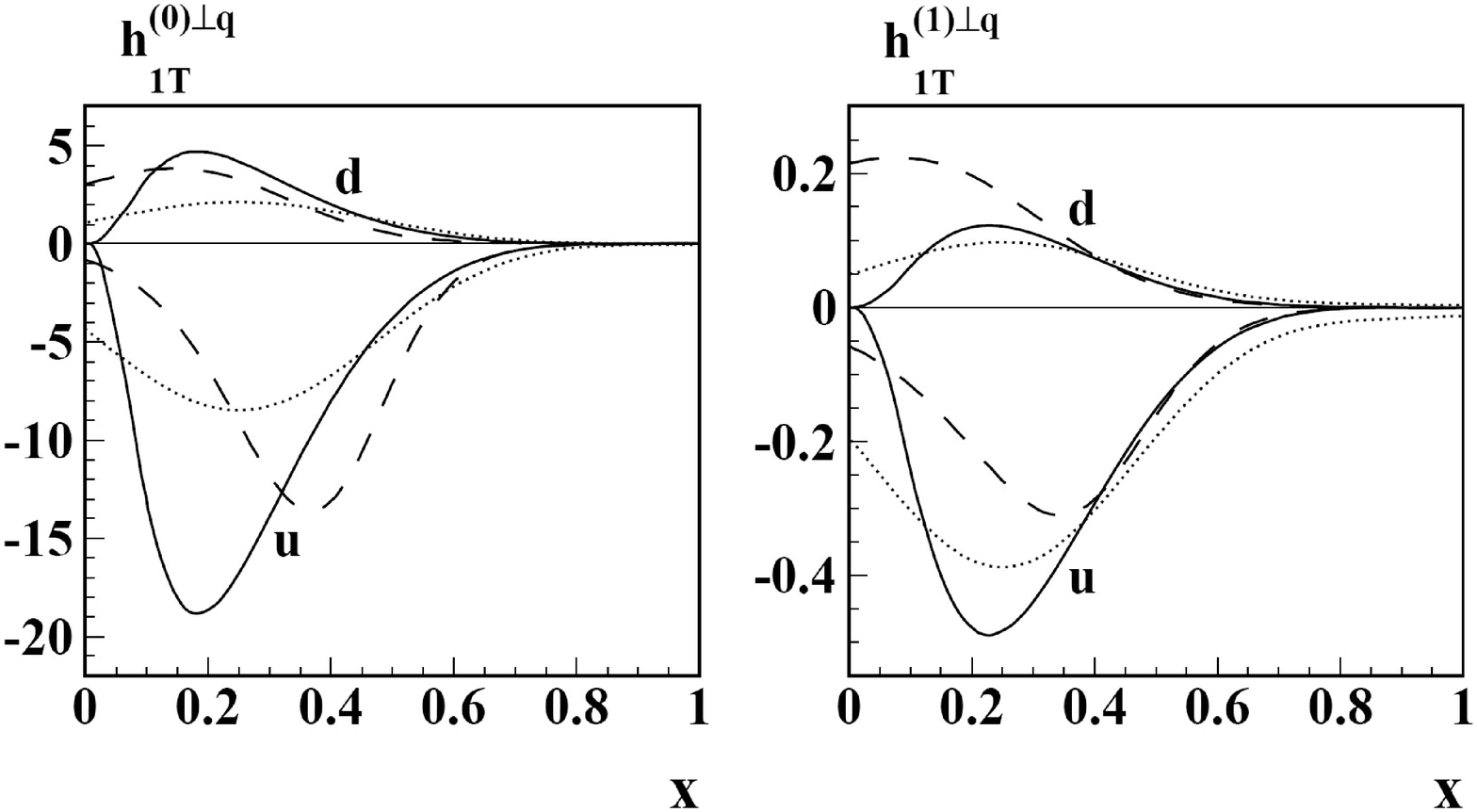}
  \caption{The transverse moments $h_{1T}^{(0)\perp\,q}(x)$  (left panel) and 
$h_{1T}^{(1)\perp q}(x)$ (right panel).
Solid curves: results from the light-cone CQM model.
Dashed curves: results from the spectator model of Ref.~\cite{Mulders3}. 
Dotted curves: results from the bag model.
}
\label{fig:h1T_int}
\end{figure}
The results for all the T-even TMDs are discussed in details in Ref.~\cite{PCB}, and here we focus only on the  distribution $h_{1T}^{\perp\,q}$.
This distribution
contributes when the quark and nucleon helicity flip in opposite directions. 
It then requires an overlap between wave function components that differ 
by two units of orbital angular momentum, either a PP or an SD interference. 
While in the case of $u$ quarks the PP and SD interference terms add with the 
same sign, in  the case of $d$ quarks they have opposite sign, 
indicating that in the present model 
the SU(6) relation between $u$ and $d$ contributions, 
$h_{1T}^{\perp\,u}= -4 h_{1T}^{\perp\,d}$, is valid for the total result 
but not for the partial wave contributions. 
The transverse moments of $h_{1T}^{\perp\,q}$
are rather different in different models, as can be seen in 
Fig.~\ref{fig:h1T_int} where the light-cone CQM results  are compared  
with results of 
the bag model and the spectator model of Ref.~\cite{Mulders3}. 
This sensitivity to the adopted model suggests that the experiments
planned at COMPASS, HERMES and JLab~\cite{Avakian-LOI-CLAS12}
could give 
useful insights to model the momentum dependence of the nucleon wave function.

According to Ref.~\cite{Miller08}, 
the distribution $h_{1T}^{\perp\,q}$ gives also a measure of the deviation
of the nucleon shape from a sphere.
This can be seen by defining a suitable spin-dependent 
quark density,  $\hat \rho_{{\rm REL}\,T}$, in a nucleon state polarized in 
the transverse direction ${\bf S}_T$ either parallel or antiparallel to 
the quark-spin direction ${\bf n}$. 
The transverse shapes of the nucleon are then 
derived from the following relation:
\begin{equation}
\frac{\hat \rho_{{\rm REL}\,T}({\bf k}_\perp,{\bf n})/M}{\tilde f_1({\bf k}_\perp^2)}
=
1 + \frac{\tilde h _1({\bf k}_\perp^2)}{\tilde f_1({\bf k}_\perp^2)}\cos\phi_n + \frac{{\bf k}_\perp^2}{2M^2}\cos(2\phi-\phi_n)\frac{\tilde h_{1T}^\perp(
{\bf k}_\perp^2)}{\tilde f_1({\bf k}_\perp^2)}, 
\label{eq:rhot}
\end{equation}
where $\phi$ is the angle between ${\bf k}_\perp$ and ${\bf S}_T$ and $\phi_n$ is the angle between ${\bf n}$ and ${\bf S}_T$. 
A tilde is placed over a given quantity to define the $x$-integrated distributions.

\begin{center}
\begin{figure}
  \includegraphics[height=7 cm]{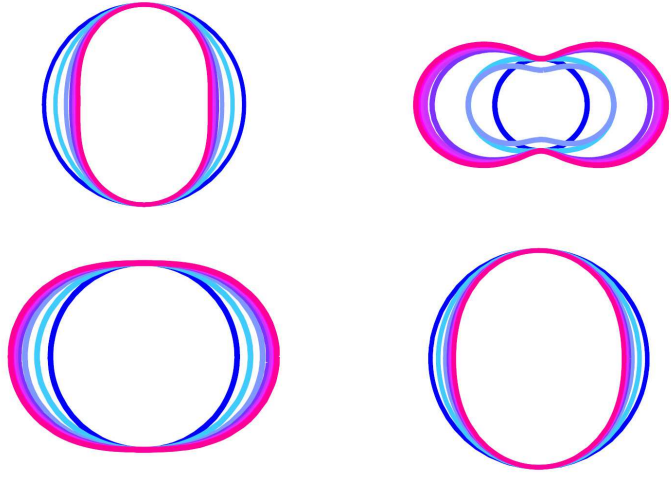}
 \caption{Transverse shape of the proton,
$\hat \rho_{{\rm REL}\,T}({\bf k}_\perp,{\bf n})/\tilde f_1({\bf k}_\perp^2)$,
assuming a struck $u$ (upper row) and $d$ (lower row) quark. 
The horizontal axis is the direction of ${\bf S}_\perp$, while  
${\bf n}=\hat{{\bf S}}_\perp$ ($\phi_n=0$) and ${\bf n}=-\hat{{\bf S}}_\perp$ ($\phi_n=\pi$)
in the left and right column, respectively.
The shapes vary from the circle to the deformed  shapes as $k_\perp$ is 
increased from 0 to 2.0 GeV in steps of 0.25 GeV. 
}
\label{fig:miller_total}
\end{figure}
\end{center}

The transverse shapes of the proton are shown in 
the Fig.~\ref{fig:miller_total}
for ${\bf S}_T$ parallel to ${\bf n}$, $\phi_n=0$ (left column), 
and for ${\bf S}_T$ antiparallel to ${\bf n}$, $\phi_n=\pi$ (right column). 
The results assuming a struck $u$ ($d$) quark are shown in  the upper (lower)
row.
 In our model $\tilde f_1^u$ ($\tilde f_1^d$) and 
$\tilde h_1^u$ (${\tilde h}_1^d$) are of the same (opposite) sign and 
similar size, so that the contribution of the first two terms on the rhs of 
Eq.~(\ref{eq:rhot}) tend to cancel each other for 
$\phi_n=\pi$ ($\phi_n=0$) emphasizing the role of the pretzelosity 
in producing deformation. For $u$ ($d$) quarks the last term in 
Eq.~(\ref{eq:rhot}) is  negative (positive) 
for $\phi=\phi_n=0$ and its size increases (reduces) with the inclusion 
of the D-wave. This explains the larger  transverse deformation in the direction antiparallel (parallel) to ${\bf S}_T$ for a struck $u$ ($d$) quark, 
with a more significant effect in the case of the $u$ quark.

\section{Spin densities in the impact parameter space}

When $\xi=0$ and $x>0$, by a two-dimensional Fourier transform to 
impact parameter space GPDs can be interpreted as densities of quarks with 
longitudinal momentum fraction $x$ and transverse location $\tvec b$ with 
respect to the nucleon center of momentum~\cite{Burkardt00a,Burkardt03}. 
Depending on the polarization of both the active quark and the parent nucleon, 
according to 
Refs.~\cite{Burkardt03,diehlhagler05} one defines three-dimensional 
densities $\rho(x,{\tvec b}, \lambda,\Lambda)$ and  $\rho(x,{\tvec b},{\tvec s}_T,{\tvec S}_T)$ representing the probability  to find a quark with longitudinal
 momentum fraction $x$ and transverse position $\tvec b$ either with 
light-cone helicity $\lambda$ ($=\pm 1$) in the nucleon with longitudinal
 polarization $\Lambda$ ($=\pm 1$) or with transverse spin $\tvec s_T$ in the
 nucleon with transverse spin $\tvec S_T$. They read
\begin{eqnarray}
\rho(x,{\tvec b}, \lambda,\Lambda) &=&  \oneh \left[ H(x,{b}^2) 
  + b^j\varepsilon^{ji} S^i_T  \frac{1}{M}\, 
       E'(x,{b}^2)
  + \lambda \Lambda \tilde{H}(x,{b}^2) \,\right] ,
 \label{eq:long}\\
\rho(x,{\tvec b},{\tvec s}_T,{\tvec S}_T) 
&= &{}\dsp \oneh\left[ H(x,{b}^2)  + s^i_TS^i_T\left( H_T(x,{b}^2)  -\frac{1}{4M^2} \Delta_b \tilde H_T(x,{b}^2) \right) \right.
\nonumber\\
& &\quad{}\dsp + \frac{b^j\varepsilon^{ji}}{M}\left(
S^i_TE'(x,{b}^2)  + s^i_T\left[ E'_T(x,{b}^2)  + 2 \tilde H'_T(x,{b}^2) \right]\right)
\nonumber\\
& &\quad\left.{}\dsp+ s^i_T(2b^ib^j - b^2\delta_{ij}) S^j_T\frac{1}{M^2} \tilde H''_T(x,{b}^2) \right] .
 \label{eq:transv}
 \end{eqnarray}
where the derivatives are defined
$
f' = \frac{\partial}{\partial b^2}\, f $, and
$
\Delta_b f
= 4\, \frac{\partial}{\partial b^2}
    \Big( b^2 \frac{\partial}{\partial b^2} \Big) f $.
In Eqs.~(\ref{eq:long})-(\ref{eq:transv}) there appear the Fourier transforms
of the GPDs for unpolarized quarks ($H$ and $E$), for longitudinally polarized quarks  ($\tilde H$ and $\tilde E$) and transversely polarized quarks
 ( $H_T$, $E_T$, $\tilde H_T$, and $\tilde E_T$).
\newline
\noindent
In Eq.~(\ref{eq:long}) the first term with $H$ describes the density of 
unpolarized quarks in the unpolarized proton. The term with $E'$ introduces
 a sideways shift in such a density when the proton is transversely polarized, 
and the term with $\tilde H$ reflects the difference in the density of quarks 
with helicity equal or opposite to the proton helicity. 
\newline
\noindent
In the three lines of Eq.~(\ref{eq:transv}) one may distinguish the three
 contributions corresponding to monopole, dipole and quadrupole structures. 
The unpolarized quark density $\oneh H$ in the monopole structure is modified
 by the chiral-odd terms with $H_T$ and $\Delta_b \tilde H_T$ when both the 
quark and the proton are transversely polarized. Responsible for the dipole 
structure is either the same chiral-even contribution with $E'$ from the 
transversely polarized proton appearing in the longitudinal spin 
distribution~(\ref{eq:long}) or the chiral-odd contribution with 
$E'_T+2\tilde H'_T$ from the transversely polarized quarks or both. 
The quadrupole term with $\tilde H''_T$ is present only when both quark
 and proton are transversely polarized.

\begin{figure}
  \includegraphics[width=12.5 cm]{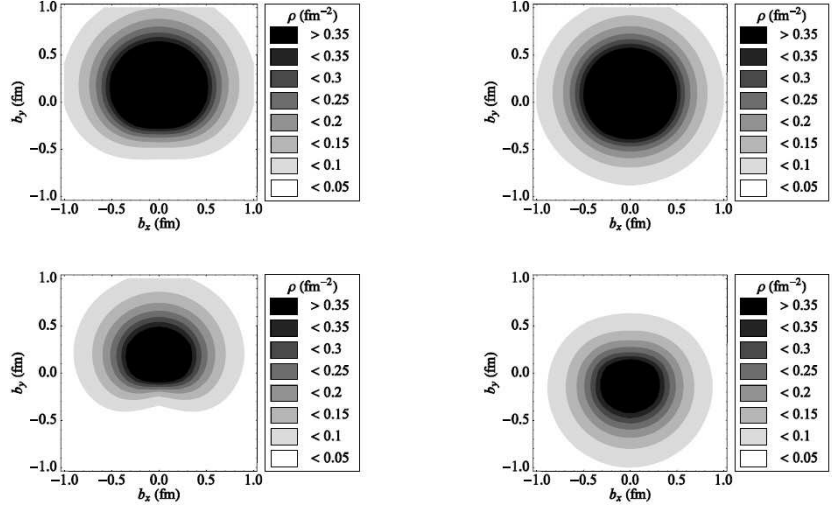}
\caption{The spin-densities
for (transversely) $\hat x$-polarized quarks in an unpolarized proton
(left column) and for unpolarized quarks in a  (transversely) $\hat x$-polarized proton (right column).
The upper (lower) row corresponds to the results for up (down) quarks.}
\label{fig:fig1}
\end{figure}

In the case of transversely polarized quarks in an unpolarized proton 
the dipole 
contribution introduces a large distortion perpendicular to both the quark 
spin and the momentum of the proton, as shown in the left column of
Fig.~\ref{fig:fig1}. 
Evidently, quarks in this situation also have a transverse component of 
orbital angular momentum. This effect has been 
related~\cite{Burkardt05b,Burkardt07} to a nonvanishing Boer-Mulders
 function~\cite{BoerMulders}  $h_1^\perp$ which describes the correlation between 
intrinsic transverse momentum and transverse spin of quarks. 
Such a distortion reflects the large value of  the anomalous tensor magnetic
 moment $\kappa_T$ for both flavors. Here, $\kappa^u_T=3.98$ 
and $\kappa^d_T=2.60$, to be compared with the values $\kappa^u_T\approx 3.0$ 
and $\kappa^d_T\approx 1.9$ of Ref.~\cite{QCDSF06a} due to a positive 
combination $E_T+2\tilde H_T$. Since $\kappa_T\sim - h_1^\perp$, the present 
results confirm the conjecture that $h_1^\perp$ is large and negative both 
for up and down quarks~\cite{Burkardt05b,Burkardt07}.

As also noticed in Refs.~\cite{Burkardt00a,QCDSF06a} the large anomalous 
magnetic moments $\kappa^{u,d}$ are responsible for the dipole distortion 
produced in the case of unpolarized quarks in transversely polarized
 nucleons (right column of Fig.~\ref{fig:fig1}). 
With the present model, $\kappa^u=1.86$ and $\kappa^d=-1.57$, 
to be compared with the values $\kappa^u=1.673$ and $\kappa^d=-2.033$ 
derived from data. This effect can serve as a dynamical explanation of
 a nonvanishing Sivers function~\cite{Sivers} $f_{1T}^\perp$ 
which measures the correlation between the intrinsic quark transverse momentum 
and the transverse nucleon spin. The present results, with the opposite 
shift of up and down quark spin distributions imply an opposite sign of 
$f_{1T}^\perp$ for up and down quarks~\cite{Burkardt02} as
 confirmed by the recent observation of the HERMES 
collaboration~\cite{Hermes05a}.
The results in Fig.~\ref{fig:fig1} are also in qualitative agreement with those obtained in lattice calculations~\cite{QCDSF06a}.

Finally,
we refer to~\cite{BPreview,PBspin}
 for the light-cone CQM results
of the densities with
more complex spin-configurations
with transverse  polarization of both the quark as well
as the proton.

\begin{theacknowledgments}
It is a pleasure for me to thank my collaborators who participated in 
different works referred to in this paper: S. Boffi, S. Cazzaniga, and M. Pincetti.            
\end{theacknowledgments}



\bibliographystyle{aipproc}   


\begin{thebibliography}{9}

\bibitem{Mulders2} 
P.J. Mulders, R.D. Tangerman, Nucl. Phys. \textbf{B 461}, 197 (1996); Erratum-\emph{ibid.} \textbf{B 484}, 538 (1997).

\bibitem{Jiarnps}
X. Ji, Ann. Rev. Nucl. Part. Sci. \textbf{54}, 413 (2004).


\bibitem{Meissner} 
S. Meissner, A. Metz, and K. Goeke, Phys. Rev. \textbf{D 76}, 034002 (2007);
K. Goeke, A. Metz, M. Schlegel, arXiv:0805.3165; arXiv:0710.5846.

\bibitem{BPP98}
S.J.~Brodsky, H.-Ch.~Pauli, S.S.~Pinsky, Phys. Rep. \textbf{301}, 299 (1998).

\bibitem{LB80}
G.P.~Lepage, S.J.~Brodsky, Phys. Rev. \textbf{D 22}, 2157 (1980).

\bibitem{Chernyak:1984ej}
V.~L. Chernyak and A.~R. Zhitnitsky,
Phys. Rep. \textbf{112}, 173 (1984).

\bibitem{King:1987wi}
I.~D. King and C.~T. Sachrajda,
  Nucl. Phys. \textbf{B 279}, 785 (1987).

\bibitem{Chernyak:1989nv}
V.~L. Chernyak, A.~A. Ogloblin,
  and I.~R. Zhitnitsky, Z. Phys.
  \textbf{C 42}, 583 (1989).

\bibitem{Braun:1999te}
V.~M. Braun, e al.,
 Nucl. Phys.
  \textbf{B 553}, 355 (1999).

\bibitem{Braun:2000kw}
V.~Braun, {\it et al.\/},
  Nucl. Phys. \textbf{B 589}, 381 (2000);
 Erratum-\emph{ibid.}\  \textbf{B 607}, 433 (2001).
 
  
\bibitem{Stefanis}
N.G. Stefanis, Eur. Phys.  J. direct \textbf{C 7}, 1 (1999).

\bibitem{BurkJiY02}
M.~Burkardt, X.~Ji, F.~Yuan, Phys. Lett. \textbf{B 545}, 345 (2002).

\bibitem{Ji:2002xn}
X.~Ji,
 J.-P. Ma  and F.~Yuan,
  Nucl. Phys. \textbf{B 652}, 383 (2003);
  Eur. Phys. J. \textbf{C 33}, 75 (2004);
 Phys. Rev. Lett. \textbf{90}, 241601 (2003).
  
\bibitem{PBff}
B. Pasquini, and S. Boffi, Phys. Rev. \textbf{D 76}, 074011 (2007);
Phys. Rev. \textbf{D 73}, 094001 (2006).

\bibitem{BPT1}
S. Boffi, B. Pasquini and M. Traini, Nucl. Phys.  \textbf{B 649}, 243 (2003);
Nucl. Phys.  \textbf{B 680}, 147 (2004).


\bibitem{PCB}
B. Pasquini, S. Cazzaniga, and S. Boffi, arXiv:0806.2298.

\bibitem{BoerMulders} 
D. Boer, P.J. Mulders, Phys. Rev. \textbf{D 57}, 5780 (1998).

\bibitem{Sivers}
D.W.~Sivers, Phys. Rev. \textbf{D 41}, 83 (1990).



\bibitem{PPB}
B. Pasquini, M. Pincetti, and S. Boffi, Phys. Rev.  \textbf{D 72}, 094029 
(2005).

\bibitem{PPB07}
B. Pasquini, M. Pincetti, and S. Boffi, Phys. Rev.  \textbf{D 76}, 034020 
(2007).
\bibitem{Avakian08}
H.~Avakian, A.~V.~Efremov, P.~Schweitzer, F.~Yuan, arXiv:0805.3355 [hep-ph].

\bibitem{Mulders3}
R. Jakob, P.J. Mulders, and J. Rodrigues, Nucl. Phys. \textbf{A 626}, 937 (1997).

\bibitem{Conti}
A. Bacchetta, F. Conti, and M. Radici,
e-Print: arXiv:0807.0323 [hep-ph].

\bibitem{Gamberg}
L.P. Gamberg, G.R. Goldstein, and M. Schlegel, Phys. Rev. \textbf{ D 77}, 094016
(2008).

\bibitem{BPreview}
S.~Boffi and B.~Pasquini, Rivista Nuovo Cim. \textbf{30}, 387 (2007).



\bibitem{Avakian-LOI-CLAS12}
  H.~Avakian,  {\it et al.}, JLab LOI 12-06-108 (2008);
JLab E05-113; JLab PR12-07-107.



\bibitem{Miller08}
G.A.~Miller, Phys. Rev. \textbf{C 76}, 065209 (2007).


\bibitem{Burkardt00a}
M.~Burkardt, Phys. Rev. \textbf{D 62}, 071503 (2000); (E) Phys. Rev. 
\textbf{D 66}, 119903 (2002).

\bibitem{Burkardt03}
M.~Burkardt, Int. J. Mod. Phys. \textbf{A 18}, 173 (2003).


\bibitem{diehlhagler05}
M.~Diehl, Ph.~H\"agler, Eur. J. Phys. \textbf{C 44}, 87 (2005).




\bibitem{Burkardt05b}
M.~Burkardt, Phys. Rev. \textbf{D 72}, 094020 (2005).

\bibitem{Burkardt07}
M.~Burkardt, B.~Hannafious, Phys. Lett, \textbf{B 568}, 130 (2008).

\bibitem{QCDSF06a}
M. G\"ockeler, {\it et al.\/}  (QCDSF/UKQCD Collaboration), 
Phys. Rev. Lett. \textbf{98}, 222001 (2007).


\bibitem{Burkardt02}
M.~Burkardt, Phys. Rev. \textbf{D 66}, 114005 (2002);
M.~Burkardt, Nucl. Phys. \textbf{A 735}, 185 (2004).

\bibitem{Hermes05a}
A.~Airapetian {\it et al.\/} (Hermes Collaboration), Phys. Rev. Lett. 
\textbf{94},
012002
(2005).

\bibitem{PBspin}
B. Pasquini, and S. Boffi, Phys. Lett.  \textbf{B 653}, 23 (2007).




\end{thebibliography}




\end{document}